\documentstyle[epsf,11pt]{skel}

\begin{document}

\newcommand{\beq} {\begin{equation}}
\newcommand{\eeq} {\end{equation}}
\newcommand{\bea} {\begin{eqnarray}}
\newcommand{\eea} {\end{eqnarray}}
\newcommand{\nn} {\nonumber}

\newcommand{\nc}{\newcommand}

\def\slash#1{\raise.18ex\hbox{/}\kern-.60em #1}

\nc{\ipfw}{I^{(W)}_{\rm pf}}
\nc{\ipaw}{I^{(W)}_{\rm pa}}
\nc{\ipfs}{I^{(S)}_{\rm pf}}
\nc{\ipas}{I^{(S)}_{\rm pa}}
\nc{\Tr}{{\rm Tr\,}}
\nc{\rome}{{\rm R}}
\nc{\ie}{{\em i.e.}}
\nc{\eg}{{\em e.g.}}
\nc{\etal}{{\em et al.}}
\nc{\calF}{{\cal F}}
\nc{\calO}{{\cal O}}
\nc{\calS}{{\cal S}}
\nc{\kapbar}{\bar{\kappa}}
\nc{\uidot}{\dot u}
\nc{\uiidot}{\ddot u}
\nc{\uiiidot}{\stackrel{\ldots}{u}}
\nc{\psii}{\psi^{(1)}}
\nc{\psibar}{\overline\psi}
\nc{\chibar}{\overline\chi}
\nc{\psiibar}{\overline\psi^{(1)}}
\nc{\psibari}{\overline\psi^{(1)}}
\nc{\epsilonbar}{\overline\epsilon}
\nc{\etabar}{\overline\eta}

\newcommand{\rnc}{\renewcommand}
\rnc{\topfraction}{1.0}
\rnc{\bottomfraction}{1.0}
\rnc{\textfraction}{0.0}

\nc{\rng}{\rangle}
\nc{\lng}{\langle}
\nc{\rcite}{Ref.\ \cite}
\nc{\ba}{\begin{array}}
\nc{\ea}{\end{array}}
\nc{\lb}{\left(}
\nc{\rb}{\right)}
\nc{\qrt}{\frac{1}{4}}

\nc{\al}{\alpha}
\nc{\bt}{\beta}
\nc{\gm}{\gamma}
\nc{\dl}{\delta}
\nc{\ep}{\epsilon}
\nc{\varep}{\varepsilon}
\nc{\zt}{\zeta}
\nc{\et}{\eta}
\nc{\th}{\theta}
\nc{\kp}{\kappa}
\nc{\lm}{\lambda}
\nc{\rh}{\rho}
\nc{\sg}{\sigma}
\nc{\ta}{\tau}
\nc{\ph}{\phi}
\nc{\vr}{\varphi}
\nc{\ch}{\chi}
\nc{\ps}{\psi}
\nc{\om}{\omega}
\nc{\noi}{\noindent}
\nc{\rr}[1]{$^{#1}$}
\nc{\rf}[1]{(\ref{#1})}
\nc{\rfs}[2]{(\ref{#1},\ref{#2})}
\nc{\smgr}{\stackrel{\textstyle <}{>}}
\nc{\grsm}{\stackrel{\textstyle >}{<}}
\nc{\aleq}{\mbox{}_{\textstyle \sim}^{\textstyle < }}
\nc{\ageq}{\mbox{}_{\textstyle \sim}^{\textstyle > }}
\nc{\ra}{\rightarrow}
\nc{\lra}{\leftrightarrow}
\nc{\be}{\begin{equation}}
\nc{\ee}{\end{equation}}
\nc{\eqrf}{Eq.\ \rf}
\nc{\erf}{{\rm erf}}

\begin{frontmatter}

\rightline{DFTUZ/95/19}
\rightline{LPTHE Orsay-95/50}
\rightline{hep-lat/9507003}
\rightline{June 1995}
\rightline{(Revised version, September 1995)}

\title{A class of chiral fermion models}

\author[Zaragoza]{ J.L.~Alonso},
\author[Paris]{ Ph.~Boucaud},
\author[Zaragoza]{ F.~Lesmes\thanksref{FELIPE}} and
\author[Zaragoza]{ A.J.~van~der~Sijs\thanksref{ARJAN}}
\address[Zaragoza]{Departamento de F\'\i sica Te\'orica,
    Universidad de Zaragoza,
     50009 Zaragoza, Spain.}

\address[Paris]{Laboratoire de Physique Th\'eorique et Hautes
Energies, Universit\'e de Paris XI, 91405 Orsay Cedex, France.\thanksref{CNRS}}
\thanks[FELIPE]{Now at ``Instituto de Magnetismo Aplicado Salvador Velayos'',
   Apdo.\ de Correos 155,\ \ 28230 Madrid, Spain.}
\thanks[ARJAN]{E-mail address: {\tt arjan@sol.unizar.es}.}
\thanks[CNRS]{Laboratoire associ\'e au CNRS.}

\begin{abstract}
We study the relation between the Roma and Zaragoza proposals
for chiral fermions on the lattice.
The fermion action in the Roma approach is shown to be equivalent to
one of the Zaragoza type.
This result is used to perform a mean-field study of the phase diagram
for chiral Yukawa models based on the Roma action.
The phase diagram is compared with the one based on the Zaragoza model
with the most local choice for the fermion interactions.
\end{abstract}
\end{frontmatter}

\parindent 1em

\section{Introduction}


\par
The formulation of a Chiral Gauge Theory (CGT) on the lattice suffers
from the well-known doubling problem   \cite{un}.
Several proposals to deal with this problem have been reviewed
in Refs.~\cite{deux,muenster}.
In some of them one tries to apply the recipe successfully used in
vector-like gauge theories:
a Wilson-like term is introduced in an attempt
to give a large mass to the unwanted doublers.
In the Smit-Swift model for example  \cite{SmitSwift},
a scalar field is introduced to write down a gauge invariant Wilson-Yukawa
term.
When the scalar field acquires a non-vanishing vacuum expectation value,
a mass term is generated for all the fermions.
In the perturbative regime in the continuum, however,
a fermion with a mass created by a Yukawa interaction does not decouple
when this mass becomes large \cite{apcar,velt,tous}. The same phenomenon
appears to occur on the lattice.
The electroweak $S$, $U$ and $\Delta\rho$ parameters, for example,
receive contributions from the doublers in the Smit-Swift model \cite{Dugan}.
In the mirror fermion model  \cite{Montvay},
supplementary, interacting physical fermions (the mirror fermions)
are introduced.
They make the model vector-like, with chiral properties at low energy,
so that a conventional gauge invariant Wilson term can be used;
the doublers get a mass
of the order of the cut-off and the parameters are tuned in such a way
that the masses of the mirror fermions are
sufficiently high to have remained unobserved so far.
These additional heavy fermions still contribute
to the above-mentioned electroweak parameters, though \cite{Dugan}.

It seems that all models which try to produce heavy masses
through the Higgs mechanism, to get rid of the unwanted fermions,
are plagued by these non-decoupling effects.
In the Zaragoza proposal \cite{ZaraCapri,ZaraMPL,ZaraPRD},
the heavy fermions are avoided.
Instead, the doublers are massless and non-interacting
(see Sect.~\ref{sec:Zara}).
In the Roma approach \cite{RomeI}, referred to as Roma I in \rcite{maiani},
auxiliary fermions are used but the decoupling of these additional particles
is achieved through a symmetry property \cite{GP}.
Both the Roma and Zaragoza approaches are gauge non-invariant regularizations.
The Roma group has shown \cite{RomeI}, however, that one can recover a
gauge invariant theory by an (in general) non-perturbative tuning of
counterterms.
The Roma and Zaragoza methods both preserve the global chiral symmetry and
it has been shown that they reproduce the correct values for the fermionic
contributions to the electroweak parameters $S$, $U$ and (the ``universal
part'' of) $\Delta\rho$ at one loop \cite{Dugan,ZaraRho}.
The variant of the Roma model discussed in \rcite{RomeII}, known as
Roma II \cite{maiani}, in which no redundant variables are used,
and the Reduced Staggered Fermion Model of Smit and
collaborators \cite{redstag},
should also correctly reproduce these parameters. However, as these
models break the global chiral symmetry one would have to take into account
the gauge non-invariant counterterms to achieve this \cite{ZaraRho}.
Recently, a lot of effort has furthermore
been dedicated to proposals related to domain-wall fermions and
fermions in the continuum \cite{shamir95}.
However, the computation
of the electroweak parameters $S$, $U$ and $\Delta \rho$ using some
implementation of these ideas has not yet been considered.

In this paper,\footnote{For a concise presentation of part of the work
described here, see Ref.~\protect\cite{Bielefeld}.} we shall establish
a direct relation between the Roma I and Zaragoza actions.
This relation is exploited to determine the phase diagram of chiral
Yukawa models based on the Roma I approach using mean-field techniques.
This phase diagram is compared with the one for the Zaragoza model with
the most local fermion interaction.

The study of such phase diagrams is important in order to investigate
where a continuum field theory may be sensibly defined.  Clearly, one
would like to recover (the gaugeless limit of) the Standard Model as a
continuum limit of a lattice model.
Furthermore, it is not excluded, that certain properties of quantum field
theories not anticipated on the basis of perturbative methods, like
non-trivial fixed points, will provide a hint how to ``improve'' or
extend the Standard Model.  The search for such hints is another
incentive to study phase diagrams of chiral Yukawa models.
In addition, the answer to questions such as: what is the maximal fermion
mass which can be generated by a Yukawa coupling, what is the effect
of a strong Yukawa coupling on the upper and lower bounds on the Higgs mass,
etc., will depend on where in the phase diagram of the, say,
SU(2)$\,\times\,$U(1)
chiral Yukawa model one approaches the continuum limit.
Most of these questions have already been studied for the Smit-Swift
model \cite{bocketaletc} and the mirror fermion model \cite{frick}.
In particular, in the strong coupling symmetric phase of the former model,
some perturbatively unexpected physics was found \cite{GPS}.
For a detailed discussion of possible continuum limits in general
see the review of Ref.\ \cite{DeJersak}.
Some of these questions are currently also under investigation
\cite{Melbourne,inprep} for the model with the local Zaragoza form factor.
Here, we will only briefly discuss some properties which
can be analysed in the context of our mean-field computations, such
as which particles survive in the continuum limit.
We would like to recall, however, that in the Roma and Zaragoza formulations
the Standard Model is defined in the scaling region on the broken side of the
small-$y$ ferromagnetic--paramagnetic phase transition.

The remainder of this paper is laid out as follows.
In Section 2, the main characteristics of the Zaragoza proposal are recalled.
Section 3 reviews the auxiliary fermion method used in the Roma I action
and establishes the equivalence with a model of the Zaragoza class.
In Section 4 we calculate and compare phase diagrams
of chiral Yukawa models in the mean-field approximation.
Some subtle aspects of the Zaragoza phase diagram, extending the
earlier work \cite{ZaraMC}, are also discussed here.
A summary and discussion can be found in Section 5.
Details of some of the calculations have been
collected in Appendices A and B.

\section {The Zaragoza proposal}
\label{sec:Zara}

The philosophy of the Zaragoza approach \cite{ZaraCapri,ZaraMPL,ZaraPRD}
(see also Ref.~\cite{ZaraRoma})
to chiral fermions on the lattice is to tolerate the presence of the species
doublers, but to prevent their communication with the real world:
they are allowed to propagate but are kept massless and non-interacting.

The kinetic term for the fermions is taken to be the naive one,
\beq
\calS_{ \rm kin} \ =\   \sum_{x,y}\; \psibar (x) K_{ x, y} \psi (y),
\label{defnaive}
\eeq
where $K$ is the usual lattice Dirac operator,
\beq
K_{ x, y} \ =\  {1\over 2} \sum_\mu
\gamma_\mu \left( \delta_{y,x+\hat\mu} \ -\
\delta_{x,y+\hat\mu}\right).
\label{defK}
\eeq
It takes care of the propagation of all $2^d$ fermion species.

\begin{figure}[p]
\centerline{
\epsfxsize=\textwidth
\epsfbox{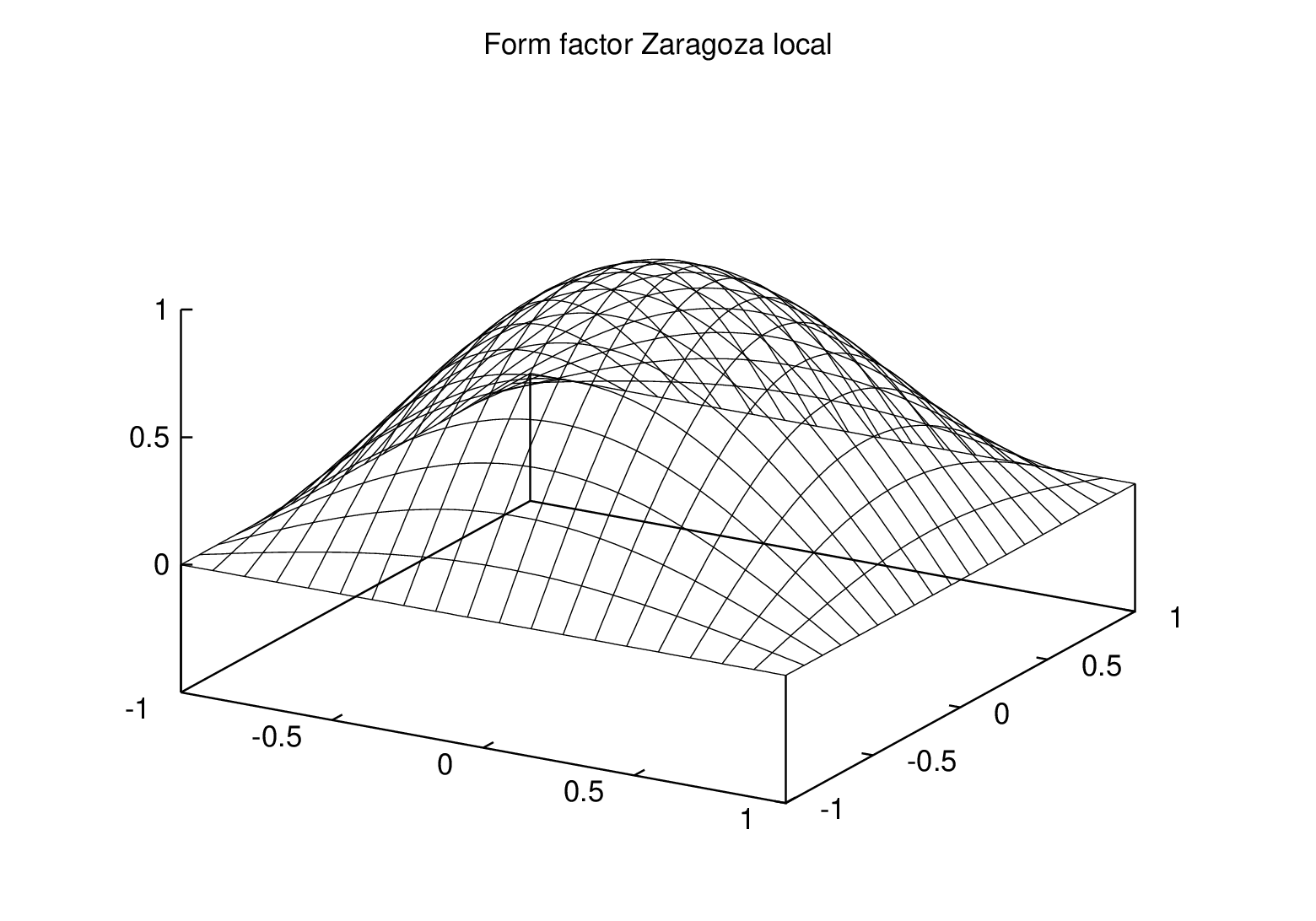}
}
\centerline{
\epsfxsize=\textwidth
\epsfbox{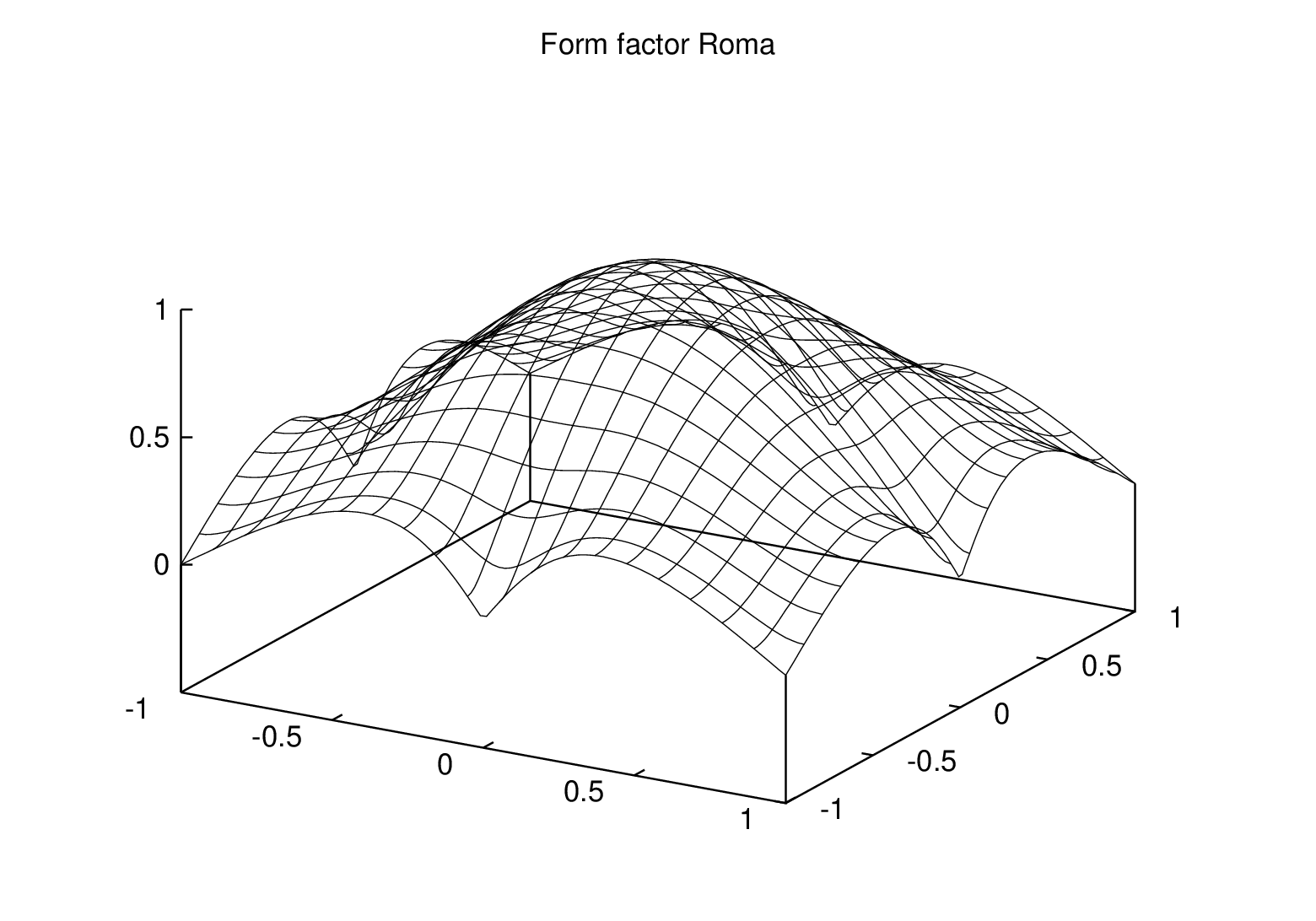}
}
\caption{
Form factors $F(p_1,p_2)$ on the Brillouin zone in two dimensions
(momenta in units of $\pi$), for the most local choice of
Zaragoza form factor \protect\rf{defF2} (top)
and for the Roma form factor \protect\rf{defFR}, with parameter $r=1$ for the
Wilson term (bottom).
}
\label{fig1}
\end{figure}

The interaction terms are to be constructed in such a way that the
doublers do not contribute.
For this purpose, we introduce the quantity $\psi^{(1)}$
defined in momentum space by
\beq
\psi^{(1)}(p) \ =\  F(p) \ \psi(p) \,.
\label{defF1}
\eeq
The ``form factor'' $F(p)$ is required to be 1 for $p = 0$, and
to vanish when $p$ equals any of the doubler momenta.
The idea is to prevent $\psii$ from coupling to the doublers at tree level.
For the function $F$, we usually choose
(a different choice was taken in Ref.~\cite{ZaraMPL}, for example)
\beq
F(p) \ =\  \prod_\mu f(p_\mu),
 \quad \quad
f(\theta) \ =\  \cos\left( {\theta \over 2 }\right),
 \quad \quad  \theta \in
(-\pi , \pi] ,
\label{defF2}
\eeq
see Fig.~\ref{fig1}.
With this form factor \rf{defF2}, the relation between the fields
$\psi^{(1)}$ and $\psi$ is the most local one in $x$-space:
$\psi^{(1)}(x)$ sits in the center of the lattice hypercube
at $x$ and is equal to the average of the $\psi$ fields
living on the corners surrounding it. This is easy to implement
in a numerical simulation.

The form factor $F$ picks out the fermion $\psii$ at $p=0$.
Similarly, one can select the fermions $\psi^{(j)}$ centered around any
of the doubler momenta $p_1 = 0$, $p_2 = (\pi,0,...)$, etc., by defining
\be
\psi^{(j)}(p) \ =\  F^{(j)}(p) \psi(p), \quad\quad F^{(j)}(p) \equiv F(p+p_j)
    \quad\quad (j=1,..,2^d)\,.
\label{defpsii}
\ee
The particular form factor of Eq.~\rf{defF2} satisfies
$\sum_j (F^{(j)})^2 = 1$, so that the kinetic term $\psibar K \psi$
reduces to $\sum_j \psibar^{(j)} K \psi^{(j)}$.
The fermionic measure in the path integral does not factorize, though.

The interaction terms are then formed in a straightforward way,
using $\psi^{(1)}$ instead of $\psi$.
For example, a lefthanded chiral gauge interaction is written as
\bea
\calS_{U\psi} &\ =\ & \half\sum_{x,\mu}
\psiibar_L(x) \gamma_\mu (U_{L\mu}(x) - 1) \psii_L(x+\hat\mu)
     \nn \\
&&\quad\quad\quad  -
\psiibar_L(x+\hat\mu) \gamma_\mu (U_{L\mu}^+(x) - 1) \psii_L(x)  \, ,
\label{LUpsi}
\eea
where, as usual, $\psii_L = P_L \psii$ with $P_L = \half (1-\gamma_5)$.
Note that this interaction involves the lattice link variable $U_{L\mu}$
with the unit matrix subtracted.
A Yukawa interaction between
$\psi^{(1)}$ and a scalar field $\Phi$ will be of the form
\beq
\calS_{\Phi\psi} \ =\  y \sum_x
\left( \psibar^{(1)}_L(x) \Phi(x) \psi^{(1)}_R(x)
     + \psibar^{(1)}_R(x) \Phi^+(x) \psi^{(1)}_L(x) \right) \, .
\label{Lyuk}
\eeq

More complicated interactions can be constructed as well.
The key point is to use only
$\psi^{(1)}$ as an interacting fermion field.
In this way, all the vertices pick up form factors in momentum space
which suppress the contributions coming from the doublers.
One is left with only one
physical, interacting fermion, while the doublers remain present
but massless and decoupled.

This decoupling  can be shown to
hold  beyond tree-level as well \cite{ZaraPRD,ZaraRoma}.
One approach is to use the Reisz power counting theorem on the lattice
\cite{Reisz} (see also Ref.~\cite{Luescher}),
which formulates (sufficient) conditions for lattice Green's functions
to tend to the corresponding continuum Green's functions (for
undoubled fermions!) in the continuum limit.
Anticipating a change of variables to be discussed shortly \rf{S2c},
we write the denominator of the propagator \rf{defpropsara}
as (temporarily restoring factors of the lattice spacing $a$)
\be
D(ap) \ =\  s^2 / (a^2 F^2) + m^2 F^2 \, ,
\label{s2}
\ee
where we have defined
\beq
s^2 \ =\  \sum_\mu \sin^2 ap_\mu \, , \qquad
m \ =\  y \langle \Phi \rangle \, ,
\label{s2m}
\eeq
and
\be
F^2 \equiv F^+ F
\label{FF}
\ee
(although we shall usually consider form factors which are
real number valued functions).
For application of Reisz' theorem, sufficient conditions are that this
denominator is a $C^\infty$ function, tending to $p^2+m^2$ in the limit
$a\ra 0$, and satisfying
\be
D(ap) \ \ge \ A\, \left(
\sum_\mu \frac4{a^2}\sin^2 (ap_\mu/2) + m^2 \right) \, ,
\label{reisz0}
\ee
on the entire Brillouin zone, for some constant $A$.
(In addition, there are a few other, rather weak conditions on the
numerators of the propagators and on the vertices which
are satisfied in the present case and, in fact, in many other cases.)
In the massless case, this inequality is easily verified for
the local form factor $F$ \rf{defF2}:
\be
s^2 / a^2 F^2 \ =\  \frac1{a^2}
    \sum_\mu \frac{4\sin^2 (ap_\mu/2) \cos^2 (ap_\mu/2)}
            {\prod_\nu \cos^2 (ap_\nu/2)}
\ \ge\  \sum_\mu \frac4{a^2}\sin^2 (ap_\mu/2) \, .
\label{reisz1}
\ee
If $m \ne 0$, we need part of the $s^2/(2a^2F^2)$ term in \rf{s2} to yield
the $A\,m^2$ term in \rf{reisz0} near the edge of the Brillouin zone, where
the factor $F^2$ multiplying the $m^2$-term becomes small.
Writing, for example,
\be
D(ap) \ =\  s^2 / (2 a^2 F^2) +  [s^2 / (2 a^2 F^2) + m^2 F^2] \, ,
\label{denom}
\ee
the term between square brackets is larger than $2^{-d}\,m^2$ in the inner
region where $p_\mu \le \pi/2$, and larger than $(1/am)^2\,m^2$ in the
rest of the Brillouin zone, and one readily finds a value of $A$ for which
the inequality \rf{reisz0} is satisfied.
The condition requiring $C^\infty$ behaviour of this
denominator is not satisfied for this local form factor, but it can
presumably be replaced by a somewhat weaker condition which would allow
application of the theorem in this case.
With a less local form factor, for instance the one following from the
Roma action (see Sect.~\ref{RalaZ}), the $C^\infty$ condition
can also be satisfied.

In addition to this perturbative decoupling theorem,
there are Golterman-Petcher shift
symmetries \cite{GP} for the doublers \cite{ZaraPRD,ZaraRoma}.
In momentum space they take the form
\bea
\psi(p) &\ \longrightarrow \ & \psi(p) + \epsilon\, \delta(p-p_i) \, , \nn\\
\psibar(p) &\ \longrightarrow \ & \psibar(p) + \epsilonbar\, \delta(p-p_i) \, ,
\label{eps}
\eea
where $\epsilon$ and $\epsilonbar$ are independent parameters,
$i=2,..,2^d$, and the $p_i$ are the doubler momenta.
On the `doubler fields' $\psi^{(j)}$ defined in Eq.~\rf{defpsii},
and their conjugates,
these symmetry transformations $(i=2,..,2^d)$ act as
\bea
\psi^{(j)}(p) &\ \longrightarrow \ & \psi^{(j)}(p) + \epsilon\delta_{ij}
    \delta(p-p_j)  \, , \nn \\
\psibar^{(j)}(p) &\ \longrightarrow \ & \psibar^{(j)}(p) +
         \epsilonbar\delta_{ij} \delta(p-p_j)  \, ,
\label{eps2}
\eea
taking care of the decoupling of the doubler with label $i$
(for more details, see Ref.~\cite{ZaraRoma}).
Clearly, the prospective physical fermion $\psii$ is not decoupled.

Evidently, the use of $\psii$ instead of $\psi$ in the interaction
terms (\ref{LUpsi},\ref{Lyuk}) does not respect the gauge invariance.
As in the Roma I model \cite{RomeI},
this means that gauge fixing and ghosts have to be introduced
explicitly in the lattice definition of the model.\footnote{Ref.\
\protect\cite{Vink} describes a proposal for a practical implementation.}
(We are not considering gauge fixing and ghost terms nor gauge-variant
counterterms here because they are not relevant for our purposes.)
A global chiral invariance is preserved, though, because $\psii(x)$ is
a linear combination of $\psi$ fields at different sites.
This global symmetry restricts the number of counterterms.
For example, no fermion mass counter-terms are needed.

As mentioned in the introduction, the Zaragoza method has been used
to study the contribution of a doublet of fermions to
the electroweak parameters $S$, $U$ and $\Delta\rho$ on the lattice.
At one loop the continuum result was reproduced \cite{ZaraRho},
confirming the decoupling of the doublers.
One should note that the gauge-variant counterterms, needed to recover
the symmetries of the continuum, do not contribute to these parameters
at the one-loop level.
A lattice simulation aiming at a non-perturbative demonstration of the
decoupling and a non-perturbative determination of the leading fermionic
contribution to the universal part of
$\Delta \rho$ is currently in progress \cite{Melbourne,inprep}.

It is interesting to note that a change of variables
\be
\psi(p) \ \ra\  F^{-1}(p)\psi(p)
\label{chofv}
\ee
transforms the original action (\ref{defnaive},\ref{LUpsi},\ref{Lyuk}),
schematically written as
\bea
\calS &\ =\ & \psibar  K \psi + \psiibar (U-K) \psii + y \psiibar \Phi \psii
\nn \\
 &\ =\ & \psibar  K \psi + \psibar F^+ (U-K) F \psi + y \psibar F^+
   \Phi F \psi
\label{S2b}
\eea
into
\be
\calS^\prime \ =\  \psibar (F^+)^{-1} K F^{-1} \psi + \psibar (U-K) \psi
   + y \psibar \Phi \psi \, ,
\label{S2c}
\ee
where $U$ is the usual gauge covariant lattice Dirac operator.
This transformation, with constant Jacobian,
shifts the form factors from the vertices to the fermion propagator.
In Feynman graphs, this can be visualized as follows.
Every fermion-fermion-Higgs vertex from \rf{S2b} contains a factor
$F^2 \equiv F^+ F$ (and similarly for the interaction with the gauge field).
These form factors can either be thought of as part of the vertex,
as in \rf{S2b}, or be assigned to the fermion propagators connected
to the vertex, as in \rf{S2c}.
In fact, in the formulation of Eq.~\rf{S2c}
the propagator (in the broken phase) has the form
\beq
S(p) \ =\  { -i \slash{s} + m F^2 \over s^2  + m^2 F^4} \ F^2 \, ,
\label{defpropsara}
\eeq
where
\beq
\slash{s} \ =\  \sum_\mu \gamma_\mu \sin p_\mu \, .
\eeq
For $p \ra 0$, \ $F\ra 1$ and the propagator corresponds to a
massive fermion.
Due to the overall factor $F^2$, however,
the poles at the doubler momenta are suppressed.

It is interesting to note that the ``Zaragoza philosophy'' mentioned
at the beginning of this Section acquires a different interpretation
in this formulation.
While in the original form \rf{S2b} species doublers were
present, albeit massless and non-interacting,
they are absent from
the $\psi$-propagator following from Eq.~\rf{S2c}.
This is not in disagreement with the no-go theorem \cite{un} because
the action \rf{S2c}, in which
the original $\psii$ is now the elementary fermion field,
contains a non-local kinetic term.

\section{The Roma I approach}
\label{sec:Roma}

One of the ideas to deal with the problem of chiral lattice fermions
has been to introduce auxiliary righthanded partners $\chi_{R}$ for the
lefthanded physical fields $\psi_{L}$.
Then a Wilson term can be formed which is used to decouple the
species doublers.
Similarly, an auxiliary lefthanded partner $\chi_L$ is introduced for
each of the physical righthanded fields $\psi_R$, with a corresponding
Wilson term.

Ideas of this kind have been discussed in
refs.~\cite{Montvay,RomeI,smitseillac,smitcapri}, with different uses.

The mirror fermion model \cite{Montvay}, in which the auxiliary fermions
are interacting fields, transforming under gauge transformations,
is essentially vectorlike in nature; the auxiliary mirror fermions are
physical.
The idea is, however, to obtain an effective chiral theory at low energies
by choosing the (both diagonal and off-diagonal) elements of the
fermion mass matrix, or the matrix of Yukawa couplings to the Higgs field,
in such a way that part of the spectrum resides at high enough energies.

If the auxiliary fields do not transform under gauge transformations,
the Wilson term breaks the symmetry.
In the work of Smit \cite{smitseillac,smitcapri}, however,
pursuing a gauge-invariant formulation,
the Higgs fields evoked by acting with gauge transformations turn
the Wilson terms into Wilson-Yukawa terms, thereby restoring the gauge
invariance.

In the Roma model \cite{RomeI}, the auxiliary fields transform only
under global transformations, so that the Wilson term breaks gauge
invariance.
The central idea of the Roma approach was, however, that the model has to
be defined in the presence of gauge-fixing and ghost terms, so one must
keep such terms also at the non-perturbative level.
At the same time, one has to include all the counterterms needed
to restore the BRST symmetry in the continuum limit.
In this formulation, the Wilson term can be naturally maintained
in its gauge non-invariant form.

We will discuss this gauge non-invariant form of the action with
auxiliary fermions, and will refer to it as the Roma I action.
In the absence of gauge fields, the mirror model action takes the
form of the Roma action if all the relevant couplings to the
mirror fields are set to zero (\ie, the bare pure and mixing mass
parameters and the Yukawa coupling for the mirror fermions).

The fermion part of the Roma I action \cite{RomeI}
consists of a minimally coupled
kinetic term and a Yukawa interaction, supplemented with a Wilson
term. It can be written in a condensed way as
\bea
\calS_{\rome} \ =\  & &
\psibar_L U_L \psi_L \ + \ \chibar_R K \chi_R \ \ + \ \
  \psibar_R U_R \psi_R \ + \ \chibar_L K \chi_L \nn\\
 & & \mbox{} + y\, \psibar_L \Phi \psi_R \ +\
               y\, \psibar_R \Phi^+ \psi_L \nn \\
 & & \mbox{} + \psibar_L W \chi_R \ + \ \psibar_R W \chi_L \ +\
               \chibar_L W \psi_R \ + \ \chibar_R W \psi_L \, .
\label{defLrome}
\eea
The physical fermion fields are denoted by $\psi$, their auxiliary chiral
partners by $\chi$.
$W$ is a `mixed' Wilson term (with parameter $r=1$),
\beq
\psibar W \chi \ =\  \
 \frac 12
    \sum_{x,\mu }\psibar(x) \, \left[ \chi(x+\hat\mu) \ +\ \chi(x-\hat\mu)
\ -\ 2 \chi(x) \right] \, ,
\label{mixwil}
\eeq
the rest of the notation is the same as in Sect.~\ref{sec:Zara}.
We have not written the kinetic terms for the gauge,
ghost and scalar fields, the scalar self-coupling, nor the
gauge-variant counter-terms, because they are of no concern here.

This action is invariant under the global chiral transformations
\bea
\psi_L \ \longrightarrow\  \Omega_L\, \psi_L & &
\chi_R \ \longrightarrow\  \Omega_L\, \chi_R \nn\\
\psibar_L \ \longrightarrow\  \psibar_L\, \Omega_L^+  & &
\chibar_R \ \longrightarrow\  \chibar_R\,\Omega_L^+  \nn\\
\psi_R \ \longrightarrow\  \Omega_R\, \psi_R & &
\chi_L \ \longrightarrow\  \Omega_R\, \chi_L \nn\\
\psibar_R \ \longrightarrow\  \psibar_R\, \Omega_R^+  & &
\chibar_L \ \longrightarrow\  \chibar_L\, \Omega_R^+
\label{omegas} \\
& \Phi \ \longrightarrow\  \Omega_L\, \Phi\, \Omega_R^+ & \nn\\
& U_{L\mu} \ \longrightarrow\
      \Omega_L\, U_{L\mu}\, \Omega_L^+ & \nn\\
& U_{R\mu} \ \longrightarrow\
      \Omega_R\, U_{R\mu}\,  \Omega_R^+ &
\nn
\eea

All the terms in $\calS_\rome$ \rf{defLrome}
containing auxiliary fields are left ungauged.
This feature distinguishes it from the mirror fermion model \cite{Montvay}.
As a consequence, the local chiral symmetry is broken.
On the other hand, one gains an invariance under
a constant shift of the $ \chi$ fields,
\bea
\chi_R(x) \ \longrightarrow \ \chi_R(x) \ +\ \eta_R \nn\\
\chi_L(x) \ \longrightarrow \ \chi_L(x) \ +\ \eta_L
\eea
from which the decoupling of these fields follows \cite{GP}.

\subsection{Roma \`a la Zaragoza}
\label{RalaZ}

Now we make the following observation.
As the auxiliary fields appear at most quadratically in
$\calS_\rome$ \rf{defLrome},
it is possible to integrate them out exactly.
This gives rise to an additional contribution
\be
-\overline\psi_L W K^{-1} W \psi_L - \overline\psi_R W K^{-1} W \psi_R
\label{deltaS}
\ee
to the kinetic term of the $\psi$ field, while the rest of the action
remains unchanged.
In the condensed notation of Eq.\ \rf{defLrome}, the action becomes
\bea
\calS^\prime_\rome & \ = \ &
\psibar_L U_L \psi_L \ - \ \psibar_L W K^{-1} W \psi_L \nn\\
&&\mbox{}\ +\ \psibar_R U_R \psi_R \ - \ \psibar_R W K^{-1} W \psi_R \nn\\
&&\mbox{}\ +\ y\, \psibar_L \Phi \psi_R \ +\ y\, \psibar_R \Phi^+ \psi_L\\
&\ =\ &
\psibar (F_R^+)^{-1} K F_R^{-1} \psi \ + \ \psibar (U-K) \psi  \nn \\
&&\mbox{}\ + \ y\, \psibar (\Phi P_R \ + \ \Phi^+ P_L) \psi,
\label{defLprime}
\eea
if we take
\beq
F_R \ =\  \sqrt{{s^2 \over s^2 + w^2}}
\label{defFR}
\eeq
($R$ for Roma), see Fig.~\ref{fig1},
with $s^2 = s^2(p) = \sum_\mu \sin^2 p_\mu$
and $w = w(p) = \sum_\mu (1  - \cos p_\mu)$.

The action \rf{defLprime} is exactly of the
Zaragoza form \rf{S2c}, with a form factor given by \rf{defFR}.

Note that $F_R(p)$ satisfies the crucial form factor requirements
$F_R(p=0) =1$ and $F_R(p=p_i) =0$ for the doubler momenta $p_i$.
A difference between the form factors \rf{defF2} and \rf{defFR}
(cf.\ Fig.~\ref{fig1}) is that $F_R=0$ only at the doubler momenta
whereas the local form factor $F$ \rf{defF2} vanishes on the
entire boundary hypersurface of the Brillouin zone.
For small momenta $p$ both form factors go as $1 - p^2/8$,
and in dimension $d=1$ they are identical.
Note that, for $d\geq 2$,
$F_R$ in \eqrf{defFR} is not well-behaved at the doubler momenta
$p_i$. It has conic behaviour, $F_R(p_i-k) \sim \sqrt{k^2}$,
causing non-differentiability at these locations.
This complicates an analytic study of the behaviour of $F_R$ in
$x$-space, needed to determine the analogue of $\psii(x)$ for this case.
Numerical calculations appear to indicate that
$F_R(x=Na,y=0)$ and $F_R(x=y=Na)$ fall off like $(-1)^N/N^3$ in
two dimensions.

It is interesting to have a look at the the fermion propagator given by
the action \rf{defLprime}.
At tree level, in the broken phase, one reads off
\bea
S(p) &\ =\   &
{ -i \slash{s} + m F_R^2 \over s^2  +
    w^2 + m^2 F_R^2}\label{defprop1} \\
            &\ =\   &
{ -i \slash{s} + m F_R^2 \over s^2  + m^2 F_R^4} \ F_R^2\label{defprop2} \; ,
\eea
where $m=y\langle \phi\rangle$.
The Wilson term $w^2(p)$ in the denominator of Eq.\ \rf{defprop1}
will decouple the doublers by the usual mechanism of giving them a
mass of the order of the cutoff,
while the conservation of global chiral symmetry is reflected by
the absence of a $w$-term in the numerator.
This is to be contrasted with the usual Wilson fermions where such a term,
although formally of order $a$,
gives a finite contribution due to its inclusion in divergent loops.

Alternatively, one may demonstrate the decoupling of the doublers
by invoking the Reisz theorem or the Golterman-Petcher shift symmetry,
as discussed in Sect.~\ref{sec:Zara}.
The denominator of the fermion propagator is now a smooth function
(\ie, it verifies the $C^\infty$-condition of Reisz' theorem),
and in the symmetric phase the required inequality follows from
\bea
s^2/a^2F_R^2 &\ =\ & (s^2 + w^2)/a^2
   \ =\  \frac1{a^2} \left\{ \sum_\mu \sin^2 ap_\mu
   + (\sum_\nu [1-\cos ap_\nu])^2 \right\} \nn \\
&\ =\ & \frac4{a^2} \left\{ \sum_\mu \sin^2 (ap_\mu/2) \cos^2 (ap_\mu/2)
+ (\sum_\nu \sin^2 (ap_\nu/2))^2 \right\}  \nn \\
&\ \ge\ & \frac4{a^2} \sum_\mu \sin^2 (ap_\mu/2) \, .
\label{reisz2}
\eea
In the broken phase, near the edge of the Brillouin zone, part of the
$w^2$-term in the denominator of the propagator \rf{defprop1} has to
provide the mass term, as in the model with the local Zaragoza
form factor, cf.\ Eq.~\rf{denom}.
In this outer region, where at least one of the momentum components
is large, $|p_\nu| \ge \pi/2$, we have $w^2 \ge 1$ and the inequality
follows easily.

The applicability of the GP shift symmetry is independent of the choice
of form factor $F$, so we simply refer to the earlier remarks
(\ref{eps}--\ref{eps2}) for the decoupling of the doublers in
Roma `\`a la Zaragoza'.

\section {Phase diagrams of Chiral Yukawa Models}

The equivalence of the Roma I action to a Zaragoza type of action
can be exploited to study the phase diagram of chiral Yukawa models
based on the Roma I action.

The phase diagram for an SU(2)$_L\,\times\,$SU(2)$_R$ Chiral Yukawa model
in the Zaragoza approach with the local form factor $F$ \rf{defF2}
was studied in Ref.\ \cite{ZaraMC},
using both mean-field techniques and lattice simulations for $n_f =2$
fermion doublets.
The mean field calculation, used in combination with a small or
large-$y$ expansion,
was shown to give a rather good approximation to the Monte Carlo
results as far as the structure of the phase diagram is concerned.

\begin{figure}
\centerline{
\epsfxsize=\textwidth
\epsfbox{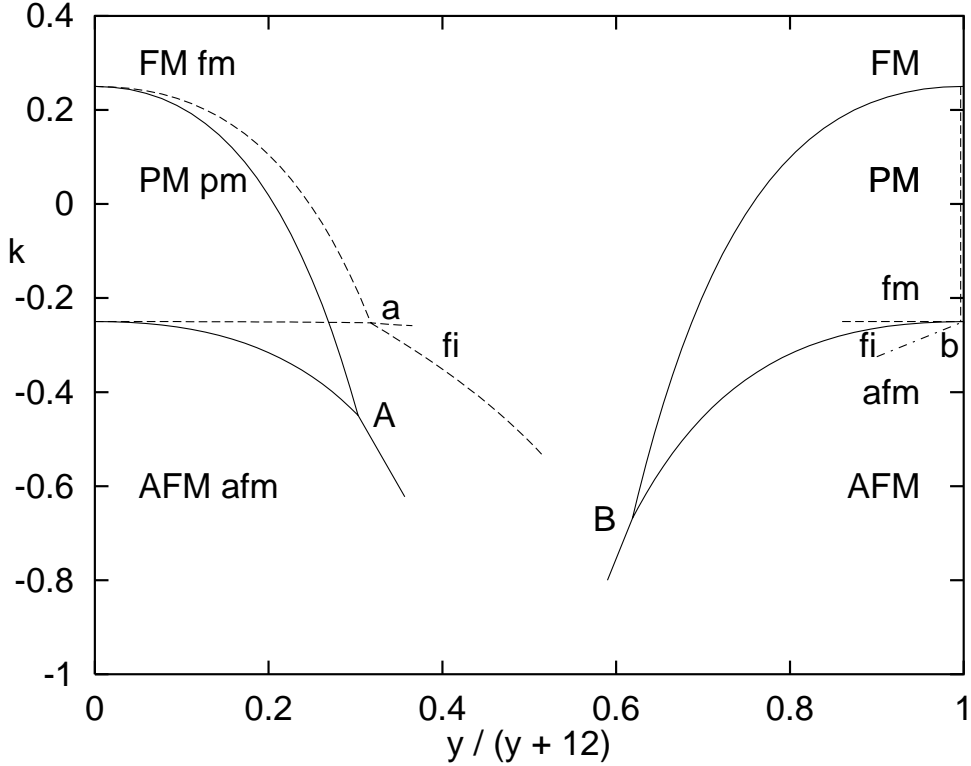}
}
\caption{
Mean-field phase diagram for SU(2)$\,\times\,$SU(2) chiral Yukawa models
with 2 fermion doublets, for the local
Zaragoza form factor (dashed curves)
and for the Roma form factor (solid curves).
Phases and multicritical points are indicated with capitals for the Roma
case and with lowercase letters for the local Zaragoza case.
The mean field result for the
large-$y$ FM--FI transition in the local case, which is
reliable only for $y \ \ageq\ 10^7$ and would be invisible for the present
choice of scale along the horizontal axis, has been extended to smaller $y$
for clarity.
The exact (mean-field) location of the FI--AFM transition
in the region of strong Yukawa coupling, for the local
form factor (dash-dotted line), is unknown.
}
\label{fig2}
\end{figure}

Here we shall study the phase diagram of the corresponding chiral
Yukawa model with the Roma I action using the mean-field techniques
of Ref.~\cite{ZaraMC}.
Some details of this approach, which combines more standard mean-field
techniques with expansions in the Yukawa coupling $y$, are described
in Appendix A.
We shall also present some new results for the Zaragoza phase diagram with
the most local form factor, extending the previous analysis \cite{ZaraMC}.
The results of the mean-field calculations
for both phase diagrams are plotted in Fig.~\ref{fig2}.

The action for $n_f$ degenerate fermion doublets is given by
\bea
\calS &\ =\ &
 -\frac k2 \sum_{x,\mu}
\Tr (\Phi^+(x) \Phi(x+\hat\mu) \ + \ \Phi^+(x+\hat\mu) \Phi(x) )
  \nn \\
&&\ +\
\sum_{i=1}^{n_f}
\left[
   \sum_{x,\mu}
   \half ( \psibar_i(x) \gamma_\mu \psi_i(x+\hat\mu) -
           \psibar_i(x+\hat\mu) \gamma_\mu \psi_i(x) ) \right.
  \nn \\
&&\ +\ \left. \sum_x y\, \psiibar_i(x) (\Phi(x)P_R +
   \Phi^+(x)P_L) \psii_i(x) \right] \, ,
\label{Scym}
\eea
for arbitrary choice of the form factor $F$ used in the definition
of the interacting fermion field $\psii$.
The modulus of the Higgs field is frozen.
This corresponds to infinite bare Higgs self-coupling $\lambda$.
With triviality in mind, we do not expect this to be a severe restriction.
The phase diagram is given in terms of the two parameters $y$ and $k$.
Because of the symmetry $\Phi\ra -\Phi$, $y\ra -y$, the $y$ values
can be restricted to the range $y\ge 0$.
For $y=0$ we have the O(4) model with critical points $\pm k_c$.
The same is true for $y=\infty$, as can be seen after rescaling the
fermion fields, $\psi\ra\psi/\sqrt{y}$. There are some subtle
aspects to the $y\ra\infty$ limit in the case of the local form factor,
though;  we will come back to this later.

In the region of weak Yukawa coupling $y$, one finds paramagnetic (PM),
ferromagnetic (FM) and antiferromagnetic (AFM) phases separated by
the second order transition lines (cf.\ Appendix A)
\bea
k^{(W)}_{\rm pf}(y) &=& {1\over 4} - y^2 { n_f\over 2}
I^{(W)}_{\rm pf}
 , \nn \\
k^{(W)}_{\rm pa}(y) &=& - {1 \over 4}- y^2 { n_f\over 2}
I^{(W)}_{\rm pa} ,
 \label{kcWpa}
\eea
where
\bea
I^{(W)}_{\rm pf} &=& \int_{-\pi}^{\pi} { {\rm d}^4 p \over (2\pi)^4 }
{ F^4(p)\over s^2(p)} ,  \nn\\
I^{(W)}_{\rm pa} &=& \int_{-\pi}^\pi { {\rm d}^4 p \over (2\pi)^4 }
{ F^2(p)F_\pi^2(p)\over s^2(p)} ,
 \label{IW}
\eea
with $F_\pi(p) \equiv F(p_1+\pi,\cdots ,p_4 + \pi)$.
These curves meet in a point $A$ with coordinates
\be
y_A \ =\  1/\sqrt{n_f\,I^{(W)}_-}  \, ,  \quad\quad
 k_A \ =\  - I^{(W)}_+ / 4 I^{(W)}_-  \, ,
 \label{ykA}
\ee
where
\be
I^{(W)}_\pm \ =\ I^{(W)}_{\rm pf} \pm I^{(W)}_{\rm pa}
 . \label{Ipm}
\ee

For the local Zaragoza form factor $F$ one finds
\be
I^{(W)}_{\rm pf} \ =\  1.6179\ 10^{-2} \, ,  \quad \quad
I^{(W)}_{\rm pa} \ =\  8.4124\ 10^{-5}
 \, , \label{Izara}
\ee
hence
\be
y_A \ =\ 7.8823/\sqrt{n_f} \, , \quad\quad
k_A \ =\ -0.25261
 \, , \label{Azara}
\ee
while for the Roma form factor $F_R$
\bea
I^{(W)}_{\rm pf} \ =\  2.5703\ 10^{-2} \, ,  \quad \quad
I^{(W)}_{\rm pa} \ =\  7.3343\ 10^{-3} \,
 , \label{Iroma}
\eea
and
\be
y_A \ =\ 7.3783/\sqrt{n_f} \, ,  \quad\quad
k_A \ =\ -0.44964
 \, . \label{Aroma}
\ee

One sees that the local form factor $F$ suppresses the
contribution from the regions near the doubler momenta stronger
than $F_R$ does:
$I^{(W)}_{\rm pa}$ is very small for the local case \rf{Iroma}
and the transition line between the
PM and AFM phases is almost horizontal.

As mentioned in Sect.~\ref{sec:Roma}, in the absence of gauge fields the
mirror fermion action takes the form of the Roma action if all the relevant
couplings to the mirrors are set to zero.
In Ref.~\cite{frick}, the mirror fermion model with SU(2)$\,\times\,$SU(2)
chiral symmetry and $n_f=2$ was simulated in essentially\footnote{A minor
difference is that the simulation of Ref.~\protect\cite{frick} was carried
out for a small non-zero bare mixing mass $\mu_{\psi\chi} = 1-8K =
\protect\calO (1/N_t^2)$.} this limit.
Our mean-field results agree quite well with Fig.~1.\ of
Ref.~\cite{frick}, showing the small-$y$ region of the phase diagram,
as can be seen after the corresponding mapping of parameters
($y \ra G_\psi/2K$ with $K\approx 0.125$).

Beyond the point $A$, one may consider the possibility of a ferrimagnetic
(FI) phase, characterized by nonzero expectation values for the fields
as well as the staggered fields (in the staggered average,
fields at the odd sites contribute with a minus sign).
The mean field method can also be applied to this situation \cite{ZaraMC},
and the equations for the transition lines can be solved numerically.
The reliability of the $y$-expansion involved, in this intermediate-$y$
region, is monitored in terms of the quantity
$|y\sum_{i=1}^{n_f}\langle\psiibar_i\psii_i\rangle|$.
In addition, one must check whether the free energy of the FI solution
is actually lower than that of the competing FM and AFM solutions.

For the local form factor, a FI phase was found \cite{ZaraMC},
separated from the FM and AFM phases by second order phase transition lines.
In Fig.~\ref{fig2}, these lines have been plotted up to the point where
$|y\sum_{i=1}^{n_f}\langle\psiibar_i\psii_i\rangle| = 1$.
Comparison with data from the lattice Monte Carlo simulation
of the model showed that the presence and location of the FI phase are
predicted quite well by the mean field results \cite{ZaraMC}.

We have used the same numerical mean field methods to look for a FI phase
in the model with the Roma form factor.
In this case, however, the assumption of second order phase transitions
between FM and FI phases,
and between AFM and FI phases, led to the inconsistent result of a FM--FI
transition line lying {\em below\/} the AFM--FI line.

\begin{figure}
\centerline{
\epsfxsize=\textwidth
\epsfbox{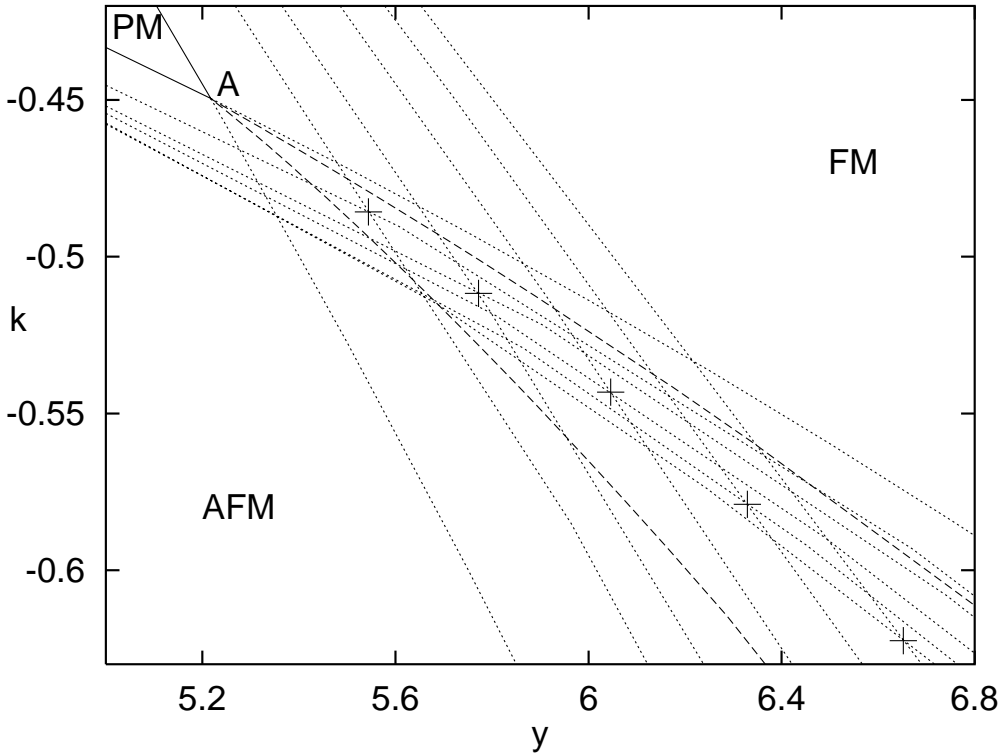}
}
\caption{
Contour plot of the free energy for FM and AFM solutions in the region
just beyond the weak-coupling PM phase.
The solid lines are the mean-field results for the second order transitions
between the PM phase, where the free energy is zero, and the FM and AFM
phases.
The mutually inconsistent (inverted) results for second order FM--FI and
AFM--FI transitions are drawn as dashed lines.
The dotted curves connect points at which FM or AFM solutions have a
given free energy.
The predominantly vertical curves are those for the FM solutions, with
free energies $0=f_0 > f_1 > \ldots > f_5$, decreasing from left to right.
The predominantly horizontal curves correspond to the AFM solutions,
with the same values $f_{0,\ldots,5}$ for the free energy, now
decreasing from top to bottom.
Points at which the FM and AFM solutions have equal free energy are
indicated with the symbol {\bf +}.
}
\label{fig3}
\end{figure}

In order to discover which of these candidate lines is (are) not
acceptable, we made contour plots of the free energy, see Fig.~\ref{fig3}.
Curves of constant free energy $\calF < 0$ were determined for FM and AFM
candidate solutions. The FM contour curves run roughly parallel to the PM--FM
transition and its continuation, where $\calF=0$, and similarly the AFM
curves parallel to the PM--AFM transition and its continuation.
Subsequently, we determined the line of intersection points, at which the
FM and AFM solutions have equal free energy.  This line was found to lie
in between the mutually inconsistent second order lines,
implying that neither of the second order lines is acceptable:
Above the line of intersection points of the free energy curves,
no AFM solution is possible because the FM solution in the same point has
lower free energy. Thus the AFM--FI line, found to lie above this
intersection line, cannot be a transition between an
AFM and a FI phase.  Similarly, the FM--FI line is unacceptable.
This is not yet sufficient to exclude a FI phase beginning at $A$.
A FI phase bounded by first order transitions would still be possible.
However, further inspection of the free energies for FM, AFM and FI mean
field solutions in this region, using similar methods,
confirmed that a FI phase is absent (at least close to the point $A$),
while the FM and AFM phases are joined by a first order transition.
This transition line is determined by the line of intersection points
in Fig.~\ref{fig3}.
It remains possible, though, that this first order line has an end point
with a FI phase opening up beyond it \cite{ferri}.

At this point it is appropriate to make a remark about the coexistence of
different phases at the point $A$.
It was stated in Ref.\ \cite{Zenkin} that, in the mean-field approximation,
the FM--PM transition line should continue beyond the point $A$ as an AFM--FI
transition, with the same slope, and similarly the AFM--PM line would
turn into a FM--FI transition line with the same slope at $A$.
The existence of the point $A$ would thus
automatically imply the presence of a FI phase.
This is not correct, as was shown in Ref.\ \cite{ferri}.
In fact, the FI phase may be absent, as in the model
with the Roma form factor studied here, and if it is present then
all four slopes at the point $A$ may be different, as in the
model with the local Zaragoza form factor.
The validity of these arguments does not depend on the details of the
mean-field approximation, such as the expansion in $y$ used to deal
with the fermion determinant.
Quantitatively, however, different values for the slopes will be found for
different approximations to the fermion determinant.
In particular, the conclusion that a FI phase is absent in the model
with the Roma form factor might have to be revised.  Apart from this,
one should of course keep in mind
the limitations of the mean-field approximation.
Fig.~1.\ of Ref.~\cite{frick}, reporting the earlier-mentioned
Monte Carlo study of the mirror
model in the limit that it is essentially equivalent to the Roma action,
displays a FI phase beyond the point $A$, but the authors
remark that this region was not investigated in detail.

Similarly,
the phase structure in the large-$y$ region can be studied by combining
mean-field with a $1/y$ expansion \cite{ZaraMC} (see also Appendix A).
One expects PM, FM and AFM phases, as in the O(4) model to which the
model reduces for $y\ra\infty$.
Assuming second order transitions between such phases, one finds
transition lines given by the curves
\bea
k^{(S)}_{\rm pf}(y) &=& {1\over 4} - \frac1{y^2} { n_f\over 2}
I^{(S)}_{\rm pf}
 , \nn \\
k^{(S)}_{\rm pa}(y) &=& - {1 \over 4}- \frac1{y^2} { n_f\over 2}
I^{(S)}_{\rm pa} ,
 \label{kcSpa}
\eea
where
\bea
I^{(S)}_{\rm pf} &=& \int_{-\pi}^{\pi} { {\rm d}^4 p \over (2\pi)^4 }
 \frac{s^2(p)}{F^4(p)} ,   \nn\\
I^{(S)}_{\rm pa} &=& \int_{-\pi}^\pi { {\rm d}^4 p \over (2\pi)^4 }
 \frac{s^2(p)}{F^2(p)F_\pi^2(p)}
 . \label{IS}
\eea
These lines meet in a point $B$ parametrized by
\be
y_B \ =\  \sqrt{n_f\,I^{(S)}_-} \, , \quad\quad
 k_B \ =\  - I^{(S)}_+ / 4 I^{(S)}_-  \, ,
 \label{ykB}
\ee
with
\be
I^{(S)}_\pm \ =\ I^{(S)}_{\rm pf} \pm I^{(S)}_{\rm pa}
 \, . \label{Ipm2}
\ee

The integrands of $I^{(S)}_{\rm pf,pa}$ \rf{IS} are just the
inverse of the integrands of $I^{(W)}_{\rm pf,pa}$ \rf{IW}.
Is is easy to show, however, that for any form factor for which the
integrals converge, both $I^{(W)}_- > 0$ and
$I^{(S)}_- > 0$, so that the points $A$ and $B$ are guaranteed to exist.
(Here we have an amusing example of positive functions $f,g$ satisfying
both $\int f > \int g$ and $\int 1/f > \int 1/g$.)

For the Roma form factor \rf{defFR}, one finds
\be
      I^{(S)}_{\rm pf} \ =\  348.01 \, ,  \quad \quad
      I^{(S)}_{\rm pa} \ =\  158.70 \,
 , \label{IromaS}
\ee
hence
\be
y_B \ =\ 13.759 \, \sqrt{n_f} \, , \quad\quad
k_B \ =\ -0.66917
  \, . \label{Broma}
\ee
As before, we also looked for a FI phase in the intermediate-$y$ region
by solving the corresponding mean field equations numerically.
The result turned out to be analogous to that for the small-$y$ equations:
An anomalous inversion between the assumed second order
FM--FI and AFM--FI transition
lines, with free energy considerations deciding in favour of FM and AFM
phases separated by a first order transition.
Again, the mean-field scenario does not exclude the possibility that
a FI phase opens up some distance away from the point $B$, while the
conclusion about the absence of a FI phase at $B$ may again be modified
by an improved treatment of the fermion determinant.

For the local Zaragoza form factor the integrals
$I^{(S)}_{\rm pf,pa}$ \rf{IS} are divergent, implying the absence of a
PM phase at large (finite) $y$ in this approximation.
FM and FI phases were found, though \cite{ZaraMC}, separated by an
almost horizontal transition line.  The $1/y$-expansion involved in
this case was estimated to be reliable for extremely large $y$ values
only, $y\ \ageq\ 10^7$.

The absence of a PM phase is reminiscent of the model of
Ref.\ \cite{leeetal330}.  There, the fermions at a site $x$
are coupled to the Higgs fields at the vertices of the hypercube at $x$.
In the present model, the Higgs field at a site $x$ is coupled to fermions
at the vertices of the hypercube at $x$, so one might be tempted by the
similarity to assign this model to the same class.
In the model of Ref.\ \cite{leeetal330}, however,
there is no equivalence to an O(4) model in the $y\ra\infty$ limit,
whereas in the Zaragoza model with the local form factor such an equivalence
is established by rescaling the fermion fields, $\psii\ra\psii/\sqrt{y}$,
after which the fermions decouple.
On the basis of this equivalence one can argue that the present model
should be of the ``funnel'' type \cite{bocketal344}, with FM, PM and
AFM phases in the large-$y$ region.

The resolution to this paradox is somewhat subtle but quite interesting.
Details of the calculations, in mean field, are given in Appendix B.
It turns out that the phase diagram of the model with the local form
factor can be considered as a limiting case of a funnel-shaped diagram,
in which the PM phase has been pushed towards $y=\infty$.  It has not
disappeared, but is contracted onto the line $y=\infty$.
The line element $(y=\infty, -k_c<k<k_c)$ is a second order
phase transition line
between the FM phase at large but finite $y$ and the PM phase {\em at\/}
$y=\infty$, and the multicritical point $B$ coincides with the
critical point $-k_c$ of the O(4) model.
In particular, the values of the mean fields are found to decay with
critical exponents (calculated explicitly in Appendix B), when this
line element is approached along lines of constant $k$.
It would be interesting to study this region, in particular
the vicinity of the point $(y=\infty,k=k_c)$,
by numerical simulation on the lattice.

An interesting question is whether the AFM phase of the O(4) model at
$y=\infty$ extends into the finite-$y$ region.   The same numerical
approach mentioned earlier is applicable in principle to investigate this.
In practice there is a technical difficulty, however.
The method makes
use of the values of certain integrals which diverge in the limit
$y\ra\infty$.  At the extremely large $y$ values required by the
reliability condition these integrals are very hard to evaluate.
We have therefore not been able to establish the precise location of the
AFM--FI transition with these methods, but there are indications that
the transition exists (the dash-dotted line in Fig.~\ref{fig2}), with the AFM
phase extending into the finite-$y$ region.

In conclusion, we have found that the SU(2)$\,\times\,$SU(2) chiral
Yukawa model \rf{Scym} with Zaragoza fermions generically has a
phase diagram of the funnel type.
An example is obtained by taking the form factor \rf{defFR}
describing the Roma action.
If the form factor is chosen to be the most local one \rf{defF2}, we
get the limiting case in which the large-$y$ PM phase is pushed towards
the line $y=\infty$.
Alternative local form factors, for which the
integrals \rf{IS} are also divergent, show similar behaviour.

In general, the dependence of the phase transition lines on the form
factor can be sketched as follows.
If the form factor $F$ suppresses the doublers strongly, \ie, if $F$
is relatively small in the vicinity of the doubler momenta,
then the integral $I^{(W)}_{\rm pa}$ \rf{IW} is very small.
The small-$y$ PM--AFM transition line will then be almost horizontal.
The integral $I^{(W)}_{\rm pf}$ \rf{IW}, which receives its dominant
contribution from momenta around the origin, will be affected only weakly,
but still the small-$y$
PM--FM transition will be curved relatively little in this case.
The PM--FM and PM--AFM transitions in the strong coupling
region, on the other hand, are determined by the integrals
$I^{(S)}_{\rm pf,pa}$ \rf{IS}, whose integrands are the inverses of
those of $I^{(W)}_{\rm pf,pa}$.
These transition lines will thus be curved relatively strongly.
As a general trend, modifying the form factor causes the transition lines
at weak coupling to bend downwards and those at strong coupling to turn
upwards, or vice versa.

\section{Discussion}
We have established a relation between the Roma I and Zaragoza methods
to put chiral fermions on the lattice.
Both proposals preserve the global chiral symmetry and have been shown
to reproduce, at one loop, the expected values of the leading fermionic
contribution to the electroweak parameters $S$, $U$ and $\Delta\rho$.
The decoupling of the fermion doublers and the protection against
mass counter-terms due to the global chiral symmetry make
these approaches well-suited for a non perturbative study
of Chiral Yukawa Theories and in particular of the
scalar-fermion sector of the Standard Model.
One can, for example, perform a lattice Monte Carlo study of the decoupling
of the doublers, and also a non-perturbative lattice computation of the most
significant contributions to $\Delta\rho$ is possible within the fermion-scalar
sector of the Standard Model \cite{Melbourne,inprep}.

We have demonstrated that the action in the Roma I formulation
(where auxiliary fermions are present) can be rewritten
`\`a la Zaragoza' with the form factor given in Eq.\ \rf{defFR}.
In $x$-space, this form factor is spread out widely,
contrary to the most local possible choice
usually made for the Zaragoza method (Eqs.\ (\ref{defF1}-\ref{defF2}))
which is easy to implement in numerical simulations.

An interesting question would be whether other globally chirally
invariant fermion actions can be rewritten in the Zaragoza
formulation, with some particular form factor.

The reformulation of the Roma I action as a Zaragoza model was
exploited to calculate the phase diagram of chiral Yukawa models based
on the Roma I action in the mean-field approximation (combined with
expansions in the Yukawa coupling $y$ or $1/y$).
The results agree quite well with the Monte Carlo simulations in the
small-$y$ region of Ref.~\cite{frick}.
In general, we argued that (mean-field) phase diagrams of chiral
Yukawa models with
Zaragoza fermions are of the ``funnel'' type, with PM, FM and AFM phases
in both the weak and strong Yukawa coupling regions.
For a class of local form factors, for which the fermion interactions
are within the range of a few lattice sites,
our mean-field calculation suggests that
the PM phase in the large-$y$ region is contracted onto the $y=\infty$
axis, part of which is at the same time a phase transition line.
One should keep in mind, however, that higher order corrections in the
$1/d$ expansion might change this conclusion.

Of course, in the scaling region the physical quantities are expected to be
the same for any form factor which kills the doublers.  In particular,
the physics at the FM side of the strong FM--PM phase transition in the
Roma case should be the same as in the model with the local form factor.
As remarked in Appendices A and B, at strong Yukawa coupling $y$ the
fermions do not survive the continuum limit, a similar conclusion as
was drawn in the Smit-Swift model.
In the small-$y$ scaling region, perturbation theory and mean field
indicate that the fermion content is as expected, with only the
physical chiral fermion surviving in the continuum limit.

Let us conclude with
a brief speculation about a different region in the phase diagram where
one might contemplate taking a continuum limit, namely the scaling
region on the AFM side of the AFM--PM phase transition.
Although the AFM and FI phases could be lattice artefacts (absent
for certain non-cubic lattice geometries),
there appears to be no rigourous argument precluding the
existence of a (possibly non-trivial) continuum limit at an AFM--PM
phase transition \cite{gallavottietc}.
If such a limit makes sense, it may be interesting to investigate,
for example, a weak-coupling scenario
with fermions which are naturally light on the scale of 246 GeV:
the fermions, with a mass proportional to $v$ in leading order would
remain (almost) massless in, say, a mean-field approach,
while the gauge boson mass squared, proportional to $k\langle \phi_x^+
\phi_{x+\hat\mu}\rangle$, would presumably not differ appreciably from
its value in the weak FM scaling region.

\vspace*{5mm}
\noindent {\large\bf Acknowledgements}

\vspace*{2mm}

This work was supported by EC contracts CHRX-CT92-0051 and ERBCHBICT941067,
by DGICYT Spain, project AEN 94-218, by Acci\'on Integrada
Hispano-Francesa HF94-150B and by the National Science Foundation
under Grant No.\ PHY94-07194.

\appendix
\section{Appendix}

Here we present some details of the mean field computation.
The most significant characteristics of the approach have been
described in \rcite{ZaraMC}.
As our starting point we shall take Eq.~11 of that reference,
giving the mean field free energy for the
SU(2)$\,\times\,$SU(2) chiral Yukawa model of Eq.~\rf{Scym},
\bea
\calF \ =\ -\frac1N \log Z & \ =\ &
\half (\lambda^2 + \bar\lambda^2) - 2dk (v^2 - \bar v^2) +
  \alpha v + \bar\alpha \bar v
\nn \\
&&\quad - \half [ u(\alpha + \bar\alpha) + u(\alpha-\bar\alpha)]
- C_0 I \, .
\label{free}
\eea
Here $C_0 = 2^{[d/2]}n_f/2$, $\,n_f$ is the number of fermion doublets,
$d$ is the dimension of space-time,
and $u(\alpha) = \log(2 I_1(\alpha)/\alpha) = \alpha^2/8 - \alpha^4/384
+ \calO(\alpha^6)$ for SU(2), with $I_1$ the modified Bessel
function of order 1.
$\calF$ is a function of the mean fields $\alpha$, $v$ and $\lambda$,
collectively denoted as $h_i$, and of the staggered mean fields
$\bar\alpha$, $\bar v$ and $\bar\lambda$, denoted as $\bar h_j$,
and is parametrized by the hopping parameter $k$ for the Higgs
field and the Yukawa coupling $y$.
The quantities $\alpha$ and $v$, and $\bar\alpha$ and $\bar v$,
are mean field values for auxiliary
fields needed to describe the scalar field, and $\lambda$ and
$\bar\lambda$ are related to
an auxiliary field corresponding to the fermion condensate
$\lng\psibari\psii\rng$ (see \rcite{ZaraMC} for details).

The function $I$ comes from the fermionic determinant.
It is obtained after a truncation of terms of higher order than
quadratic in $y$ (or $1/y$) in the mean field average of the Yukawa
terms.
In fact, the real expansion parameter here is
$yn_f\langle \psiibar\psii\rangle$ rather than $y$ (and
$y^{-1}n_f\langle \etabar^{(1)}\eta^{(1)}\rangle$ rather than $y^{-1}$ in the
large-$y$ region, where $\eta$ and $\etabar$ are auxiliary fermion fields
\cite{ZaraMC}), so that this approximation should be valid in and
near the paramagnetic phases \cite{ZaraMC}.
For illustration we shall consider the region of weak Yukawa coupling,
where $I$ has the form
\be
I = \int_{-\pi}^{\pi} \! \frac{d^dp}{(2\pi)^d} \; \log
\frac{[s^2(p) + (m^2 - \bar m^2)F^2(p) F_\pi^2(p)]^2
       + m^2 s^2(p) [F^2(p) - F_\pi^2(p)]^2}{s^4(p)} \, .
\label{a02}
\ee
Here $F(p)$ is the form factor, $F_\pi(p) = F(p_1+\pi,\ldots,p_d+\pi)$,
$s^2(p) = \sum_\mu \sin^2(p_\mu)$,
and $m=(m_+ + m_-)/2$, $\,\bar m = (m_+ - m_-)/2$ with
\bea
m_\pm &\ =\ & y \, \left\{ \uidot(\alpha \pm \bar\alpha) -
    (\lambda \pm \bar\lambda) \sqrt{\uiidot(\alpha \pm \bar\alpha)}
   \right\} \, .
\label{mpm}
\eea
The dots on the function $u(\alpha)$ denote derivatives.
We have normalized \eqrf{a02}
such that the free energy of the PM phase is zero.
Concerning the mass $m_f$ (in lattice units) of the physical fermions
in the scaling region of the FM phase, we can, as usual, distinguish a
weak and a strong FM region.
In fact, it can be shown from Eqs.\ (\ref{free},\ref{a02}) that $m_f^2
= m_+^2$ ($=y^2\,z^2$ in the notation of Ref.\ \cite{ZaraMC})
in the small-$y$ FM phase.  For large $y$ the result is $m_f^2 = y^2/z^2$
\cite{ZaraMC}.

To find the phase of the system for given $y$ and $k$,
the free energy is minimized with respect to the mean fields
(for a recent concise discussion see Ref.~\cite{ferri}),
for which a necessary condition is that the saddle-point equations
\be
\frac{\partial\calF}{\partial h_i} \ =\
\frac{\partial\calF}{\partial \bar h_j} \ =\ 0 \, .
\label{dF}
\ee
are satisfied.
A second order phase transition from a symmetric to a broken phase occurs
when a negative mode develops and the mean fields acquire non-zero values.
In general, such a transition is given by the condition
\be
\det \calF'' (h,\bar h) \ =\ 0
\, .
\label{d2F}
\ee

This equation is hard to solve for the case that
there are broken phases on both sides of the transition under
consideration (for example, a transition between
antiferromagnetic (AM) and ferrimagnetic (FI) phases).
Then Eq.~\rf{d2F} has to be satisfied for the non-trivial mean field values
given by Eq.~\rf{dF}.
If one wants to solve these equations perturbatively
by expanding $\calF$ in powers of the mean fields,
then it is crucial \cite{ferri} to include terms up to
at least fourth order in the fields.
Alternatively, a numerical approach may be adopted to find the
exact solution, as in Ref.~\cite{ZaraMC} and in the present paper.
It should be emphasized, however, that the assumption of second order
phase transition lines should be checked afterwards by comparing free
energies of different mean field values.

If either side of the transition is PM, however, then
Eq.~\rf{d2F} is taken at zero mean fields and can be solved easily.
One can for example expand $\calF$ to quadratic order in the fields,
\be
\calF \ =\  h^+ M h \ + \ \bar h^+ \bar M \bar h
 \ +\ \calO\left((h,\bar h)^4\right) \, ,
\label{free2}
\ee
where
\be
\!\!\!\!\! h \ =\
 \left( \begin{array}{c} \alpha \\ v \\ \lambda \end{array} \right)
\, , \quad\quad
M\ =\
 \left( \begin{array}{ccc}
           -\frac18-\frac1{8}C_0 y^2 \ipfw & \half & \frac14 C_0 y^2 \ipfw \\
           \half   &   -2dk   &   0   \\
           \frac14 C_0 y^2 \ipfw  & 0 &  \half - \half C_0 y^2 \ipfw
        \end{array}   \right)
\, ,
\label{hm1}
\ee
\be
\!\!\!\!\! \bar h \ =\
 \left( \begin{array}{c} \bar \alpha \\ \bar v \\ \bar \lambda
          \end{array} \right)
\, , \quad\quad
\bar M\ =\
 \left( \begin{array}{ccc}
             -\frac18+\frac1{8}C_0 y^2 \ipaw & \half & -\frac14 C_0 y^2 \ipaw
\\
             \half   &   2dk   &   0   \\
             -\frac14 C_0 y^2 \ipaw  & 0 &  \half + \half C_0 y^2 \ipaw
        \end{array}   \right)
\, ,
\label{hm2}
\ee
with
\be
\ipfw \ =\  \int_{-\pi}^\pi { {\rm d}^d p \over (2\pi)^d }
{ F^4(p)\over s^2(p)} \, , \quad\quad\quad
\ipaw \ =\  \int_{-\pi}^\pi { {\rm d}^d p \over (2\pi)^d }
{ F^2(p)F_\pi^2(p)\over s^2(p)} \, .
 \label{IWapp}
\ee
 From Eqs.~(\ref{dF},\ref{d2F}) one finds the curves
\bea
k^{(W)}_{\rm pf}(y) &\ =\ & {1\over d} \left( 1 - \frac{2^{[d/2]}}{2}
    n_f y^2 \ipfw \right) \,
 , \nn \\
k^{(W)}_{\rm pa}(y) &\ =\ & {1 \over d} \left( - 1 - \frac{2^{[d/2]}}{2}
    n_f y^2 \ipaw \right) \, ,
 \label{kcWpaapp}
\eea
as second order transition lines for the PM--FM and PM--AM transitions
in the small-$y$ region, respectively.
For $d=4$ they coincide with the formulas \rf{kcWpa}
in the main text, and for $y=0$ one recovers the phase structure of the
O(4) model.

The region of strong Yukawa coupling $y$ can be studied similarly, with
the integral $I$ of Eq.~\rf{a02} replaced by its large-$y$
analogue (see Ref.~\cite{ZaraMC}).
One finds PM, FM and AM phases separated by the lines
\bea
k^{(S)}_{\rm pf}(y) &\ =\ & {1\over d} \left( 1- \frac{2^{[d/2]}}{2}
    n_f \frac1{y^2} \ipfs \right) \,
 , \nn \\
k^{(S)}_{\rm pa}(y) &\ =\ & {1\over d} \left( -1- \frac{2^{[d/2]}}{2}
    n_f \frac1{y^2} \ipas \right) \, ,
 \label{kcSpaapp}
\eea
with
\be
\ipfs \ =\  \int_{-\pi}^\pi { {\rm d}^d p \over (2\pi)^d }
{ s^2(p)\over F^4(p)} \, , \quad\quad\quad
\ipas \ =\  \int_{-\pi}^\pi { {\rm d}^d p \over (2\pi)^d }
{ s^2(p)\over F^2(p)F_\pi^2(p)} \, ,
 \label{ISapp}
\ee
{\em provided these integrals converge.}
For the Roma form factor \rf{defFR} they do converge in $d=4$ dimensions.
For the local Zaragoza form factor \rf{defF2}, however,
they are divergent for $d=4$, suggesting that large-$y$ PM and AM
phases are absent in this model \cite{ZaraMC}.
In this case there are some subtle aspects to the phase structure
in the limit $y\ra\infty$, however, which will be discussed separately
in Appendix B.

\section{Appendix}

As we have seen, the mean field determination of the phase structure of the
SU(2)$\,\times\,$SU(2) chiral Yukawa model with the local form factor
\rf{defF2}
\be
F(p) \ =\ \prod_{\mu=1}^d \cos (p_\mu/2)
\, ,
\label{defF2a}
\ee
in the strong Yukawa coupling region, is complicated by
the divergence of certain integrals \rf{ISapp}.
It was concluded in Ref.~\cite{ZaraMC} that there are no PM and AM phases
at large $y$, in contrast with the O(4) model to which
the model should reduce in the limit $y\ra\infty$, cf.\ the discussion
in Sect.~4.
Here we analyse this issue carefully and resolve the paradox.

First we investigate, in the large-$y$ region along the line
$\kappa=0$, whether or not there is a FM solution to the saddle-point
equations with lower free energy than the PM candidate solution
given by $\Lambda = v = \alpha =0$.
As we will see, the answer depends crucially on the form factor $F$.
For the local form factor \rf{defF2a}, there turns out to be a
FM solution which is favoured relative to the PM solution
for all large but finite $y$,
as suggested by the (minus) infinite coefficient of the $1/y^2$ terms
in Eq.~\rf{kcSpaapp}.

We start from Eq.~\rf{free} for the mean field free energy $\calF$ and take
the staggered fields equal to zero.
The saddle-point equations to be satisfied by the mean fields are
\bea
\alpha &\ =\ & 4d k v \, , \nn \\
v &\ =\ & \uidot + \Lambda + \frac{\Lambda^2 \uiiidot}{2\uiidot^2} \, ,
\label{a03} \\
\Lambda &\ =\ & 4 C_0 \uiidot \int_{-\pi}^{\pi} \! \frac{d^dp}{(2\pi)^d} \;
\frac{(\uidot+\Lambda) s^2}{(\uidot+\Lambda)^2 s^2 + y^2 F^4} \, , \nn
\eea
where $\Lambda = -\lambda\sqrt{\uiidot(\alpha)}$, and
$C_0 = 2^{[d/2]}n_f/2$.

For $k=0$,
\be
\calF \ =\  2 \Lambda^2 - 2 C_0 \int_{-\pi}^{\pi} \!
\frac{d^dp}{(2\pi)^d} \; \log \frac{\Lambda^2 s^2 + y^2 F^4}{y^2 F^4} \, ,
\label{a04}
\ee
and Eq.~\rf{a03} simplifies to
\bea
\alpha &\ =\ & 0 \, , \nn \\
v &\ =\ & \Lambda \ =\ -\half \lambda \, ,
\label{a05} \\
1 &\ =\ & C_0 \frac1{y^2} I_0(v/y)
 \, , \nn
\eea
where the function
\be
I_0 (x)\ =\
\int_{-\pi}^{\pi} \! \frac{d^dp}{(2\pi)^d} \; \frac{s^2}{x^2 s^2 + F^4}
\label{I0}
\ee
is seen to increase with decreasing $x$.

For not too large $y$, these equations \rf{a05}
have a non-trivial solution, and using the inequality
\be
\forall q > 0:\ \ \ \
\frac{q}{q+1} \ < \   \ln(1+q) \, ,
\label{a09}
\ee
with
\be
q \ =\  \frac{\Lambda^2 s^2}{y^2 F^4} \, ,
\label{a10}
\ee
one can verify that any such solution has lower
free energy $\calF$ than the competing PM solution.
This holds for both the Roma form factor $F_R$ \rf{defFR} and the
local form factor \rf{defF2a}.

Whether there is a FM solution for arbitrarily large $y$ depends on
the small-$x$ behaviour of the function \rf{I0}.
If it does not blow up for $x\ra 0$, \ie, if $I_0(0)$ is a convergent
integral,
then $I_0(v/y)$ is bounded from above by $I_0(0)$ and
the last equation of \rf{a05} admits no non-trivial solution
for large enough $y$,
implying a transition to a PM phase at a finite value of $y$.
This happens for the Roma form factor $F_R$ \rf{defFR} in four dimensions.

For the local form factor \rf{defF2a}, however,
the fact that $I_0(x)$ blows up for $x\ra 0$ means that for all $y<\infty$
there is a FM solution $v$ to this equation.
In fact, we will demonstrate that
\be
I_0(x) \ \sim\  x^{-3/2} \, \left(\log \frac1x\right)^{d-2}
        \quad \quad \quad\quad (x\ra 0) \, ,
\label{I0x}
\ee
in dimensions $d\ge 2$ (and $\sim x^{-1}$ for $d=1$).
This implies that the FM solution to Eq.~\rf{a05} behaves like
\be
v \ =\  \Lambda \ \sim\  \left(\frac1y\right)^{1/3}
   \, \left(\log y\right)^{(2/3)(d-2)}
\quad\quad \quad\quad \quad\quad
(y\ra\infty) \,
\label{vy}
\ee
(we shall assume $d\ge 2$ from now on),
indicating an approach to a PM phase with a second order
transition \ {\em at\/} \ $y=\infty$,
with critical exponents given by \rf{vy}.

This argument can be generalized for the approach of the $y\ra\infty$ limit
along any line of constant $k$ satisfying $-k_c \le k \le k_c$,
where $k_c=1/d$ is the O(4) critical point.
Since we are close to a PM phase, we can expand Eq.~\rf{a03} to
lowest order in the mean fields, provided $k_c-k$ is not too small.
Using again the small-$x$ behaviour \rf{I0x} of $I_0(x)$, we find that
\bea
v\ &\ \sim\ &\
         \left(\frac{k_c}{k_c-k}\right)^{2/3} \, \left(\frac1y\right)^{1/3}
   \, \left|\log \left(y\sqrt{\frac{k_c-k}{k_c}}\right)\right|^{(2/3)(d-2)}
\, , \label{vyk1} \\
\alpha\ &\ =\ &\
   4 \frac{k}{k_c} \, v \, , \qquad\qquad
       \Lambda\ \sim\ \left(\frac{k_c-k}{k_c}\right) \, v
\, , \label{vyk2}
\eea
for $y\ra\infty$ along lines of constant $k<k_c$.
The width of the small-$(1/y)$ region where this approximation
is valid shrinks to zero for $k\ra k_c$, and for $k=k_c$ we find instead:
\bea
v\ &\ \sim\ &\ \alpha\ \sim\
         \left(\frac1y\right)^{1/7}
   \, \left(\log y\right)^{(2/7)(d-2)}
\, , \label{vyk1aa} \\
\Lambda\ &\ \sim\ &\ v^3 \ \sim\
        \left(\frac1y\right)^{3/7}
   \, \left(\log y\right)^{(6/7)(d-2)}
\, . \label{vyk2aa}
\eea

When the point $(y=\infty,k=k_c)$ is approached from below along
curves of the form $1/y \sim ((k_c-k)/k_c)^\beta$ ($\beta > 0$),
the mean fields go like
\bea
v\ &\ \sim\ &\ \alpha\ \sim\
(k_c-k)^{(\beta-2)/3}  \, ,
\quad
\Lambda \ \sim \
(k_c-k)^{(\beta+1)/3}
\quad\ (\beta \ge 7/2) \, ,
\\
v\ &\ \sim\ &\ \alpha\ \sim\
(k_c-k)^{\beta/7}  \, ,
\quad \ \,
\Lambda \ \sim \
(k_c-k)^{3\beta/7}
\quad (0 \le \beta \le 7/2) \, ,
\eea
up to logarithmic corrections.
In the limiting case $\beta\ra\infty$, corresponding to approaching the
point from below on the $y=\infty$ axis, one finds
$v\sim\alpha\sim\Lambda\sim 0$, as expected for
the O(4) model in the PM phase.

We see that in this scaling region the fermion mass $m_f^2 \sim y^2/v^2$ blows
up when the phase transition is approached, so that the fermions do not
survive the continuum limit.

For $k> k_c$ one finds in the limit $y\ra\infty$ along lines of constant $k$,
for $k-k_c$ not too small:
\bea
\alpha\ &\ \sim\ &\ \left(\frac{k-k_c}{k_c}\right)^{1/2}  \, ,
\quad\quad\quad\quad
v\ =\ \frac{k_c}k \frac\alpha4 \, ,
\\
\Lambda \ &\ \sim \ & \
\left(\frac{k-k_c}{k_c}\right)^{-1/4} \left(\frac1y\right)^{1/2}
\, \left| \log \left(y\sqrt{\frac{k_c}{k-k_c}}\right)\right|^{d-2}
\, ,
\eea
which reproduces the mean field results for
the FM phase of the O(4) model for $y=\infty$.

Note finally that in all cases considered here
the asymptotic behaviour of the product of
mean fields $\alpha\Lambda^2$ or $v\Lambda^2$ is
$\sim (1/y)|\log(y/v)|^{2(d-2)}$, as can also be deduced directly
from Eq.\ \rf{a03}.

In conclusion, we see that,
for the local form factor \rf{defF2},
the vertical line element $y=\infty,
k\in[-k_c,k_c]$ is a second order phase transition line connecting
the large-$y$ FM phase in the region $y<\infty$
to the PM phase of the O(4) model at $y=\infty$.
This result also implies that the divergence of the integral $\ipas$
\rf{ISapp} becomes meaningless, because the transition line
$k_{pa}^{(S)}(y)$ between PM and AM phases has shrunk to a single point
$B = (y=\infty,k=-k_c)$ on the $y=\infty$ axis.

The promised demonstration concerning the $x\ra 0$
behaviour \rf{I0x} of the integral $I_0(x)$
for the local form factor \rf{defF2a} is as follows.
We shall consider the function
\be
I_1(x) \ =\
x^2 I_0(x) \ =\
\int_{-\pi}^{\pi} \! \frac{d^dp}{(2\pi)^d} \; \frac{x^2 s^2}{x^2 s^2 + F^4}
\ =\
\int_{-\pi}^{\pi} \! \frac{d^dp}{(2\pi)^d} \; \frac{x^2 s^2}{x^2 s^2 + F_\pi^4}
\, .
\label{I1}
\ee

For $d=1$ this integral can be computed exactly.
One finds
\be
I_1(x) \ =\
\int_{-\pi}^{\pi} \! \frac{dp}{2\pi} \;
  \left( 1 + \frac{\cos^2(p/2)}{(4x^2)\sin^2(p/2)} \right)^{-1}
  \ =\
\frac{2x}{1+2x} \, .
\label{I1a}
\ee

For the general case $d\ge 2$ we shall show that,
for small enough $x$, $I_1(x)$ satisfies
\be
C_1 \, x^{1/2} \left(\log \frac1x\right)^{d-2} \ < \   I_1(x) \ < \
C_2 \, x^{1/2} \left(\log \frac1x\right)^{d-2} \, ,
\label{I1b}
\ee
for positive constants $C_{1,2}$.

A simple order of magnitude
estimate already hints at an $x^{1/2}$ behaviour.
The dominant contribution to the integral $I_1(x)$ for small $x$
comes from the regions where $F^4 \ \aleq\  x^2s^2$, such that the
integrand is $\calO(1)$.
The $x$-dependent size of these regions determines how large this
contribution is.
In this way one finds that the dominant contribution
is of order $x^{1/2}$ and comes from the
regions where $N$ ($1\le N\le d-1$) of the momentum components $p_\mu$
are close to $\pi$, with the other $p_\mu$ away from both $0$ and $\pi$.

Rigorous bounds on $I_1(x)$ are obtained as follows.
For the lower bound one has $s^2(p) \ge \sin^2 p_1$, so that
\be
I_1(x) \ \ge \
\int_{-\pi}^{\pi} \! \frac{d^dp}{(2\pi)^d} \;
  \left( 1 + \frac{\cos^2(p_1/2)}{(4x^2)\sin^2(p_1/2)}\prod_{\mu=2}^d
          \sin^4(p_\mu/2) \right)^{-1}
\, .
\label{I1aa}
\ee
The integration over $p_1$ can be done using the $d=1$ result \rf{I1a},
and subsequently the $p_2$-integration gives
\be
I_1(x) \ \ge \
\int_{-\pi}^{\pi} \! \frac{dp_3}{2\pi} \ldots \frac{dp_d}{2\pi}\;
\sqrt{\frac{2x}{2x+\sin^2(p_3/2)\ldots \sin^2(p_d/2)}}
\ .
\label{I1ab}
\ee

For the upper bound, we divide the integration region into subregions
$R_j$ defined such that $p\in R_j$ if $\sin^2 p_j\ge\sin^2 p_k$ for
all $k\neq j$:\ \
$\int_{-\pi}^{\pi} \! d^dp/(2\pi)^d = \sum_j \int_{R_j}$.
On $R_j$, $s^2 \le d\sin^2 p_j$, and using symmetry and positivity of
the integrand one obtains
\bea
I_1(x) &\ \le \ &
\sum_j \int_{R_j} \!
  \left( 1 + \frac{\cos^2(p_j/2)}{(4dx^2)\sin^2(p_j/2)}\prod_{\mu\neq j}
          \sin^4(p_\mu/2) \right)^{-1}
\nn\\
&\ \le \ &
d\, \int_{-\pi}^{\pi} \! \frac{d^dp}{(2\pi)^d} \;
  \left( 1 + \frac{\cos^2(p_1/2)}{(4dx^2)\sin^2(p_1/2)}\prod_{\mu=2}^d
          \sin^4(p_\mu/2) \right)^{-1}
\, ,
\label{I1ac}
\eea
leading to an expression similar to \rf{I1ab}.

Using furthermore that $(p_k/\pi)^2 \le \sin^2(p_k/2) \le (p_k/2)^2$,
we get
\be
G\left[ x \left(\frac2\pi\right)^{2(d-2)} \right]
\ \le\
I_1(x)
\ \le\
d\,G\left[ x \sqrt{d}\right] \, ,
\label{I1ad}
\ee
where
\be
G[x] \ \equiv\
\int_0^1 \! dk_3\ldots \int_0^1 \! dk_d \;
      \sqrt{\frac{2x}{2x+k_3^2\ldots k_d^2}}
\ .
\label{I1ae}
\ee
We see that the limiting behaviour of $I_1(x)$ is determined by that of
$G[x]$.

In order to obtain an upperbound on $G[x]$ we divide the integration
region into $2^d$ subregions $S_i$ by writing
\be
\int_0^1 \! dk_\mu
  \ =\
\int_0^{\; \sqrt{2x}} \! dk_\mu  \ +\ \int_{\sqrt{2x}}^1 \! dk_\mu
 \quad \quad \quad \quad  (\mu=3,\ldots,d) \ .
\ee
On those regions $S_i$ where one or more of the $k_\mu$ are smaller
then $\sqrt{2x}$, we write
\be
I_{S_i} \ =\ \int_{S_i}
\sqrt{\frac{2x}{2x+k_3^2\ldots k_d^2}}
 \ \le \
\int_{S_i} 1 \ \le\ \sqrt{2x}
\, .
\ee
The remaining contribution is
\bea
&&
\int_{\sqrt{2x}}^1 \! dk_3 \ldots \int_{\sqrt{2x}}^1 \! dk_d  \;
\sqrt{\frac{2x}{2x+k_3^2\ldots k_d^2}}
\nn\\
&&
\quad \quad \ \le\
\int_{\sqrt{2x}}^1 \! dk_3 \ldots \int_{\sqrt{2x}}^1 \! dk_d  \;
\frac{\sqrt{2x}}{k_3\ldots k_d}
\ =\
\sqrt{2x}\,\left( \frac12 \log\frac1x \right)^{d-2}
\, .
\label{IRj}
\eea

To obtain a lower bound on $G[x]$, we write
\be
G[x] \ \ge\
\int_{(2x)^\beta }^1 \! dk_3 \ldots \int_{(2x)^\beta }^1 \! dk_d \;
\sqrt{\frac{2x}{2x+k_3^2\ldots k_d^2}}
\, ,
\label{I1k}
\ee
guided by what we have seen is a momentum region providing a
large contribution to the integral.
Taking $\beta = 1/(2d-4)$, we get
\be
\!\!\!\!
G[x] \ \ge\
\int_{(2x)^\beta }^1 \! dk_3 \ldots \int_{(2x)^\beta }^1 \! dk_d \;
\frac{\sqrt{x}}{k_3\ldots k_d}
\ =\
\sqrt{x}\,\left( \frac1{2(d-2)} \log\frac1x \right)^{d-2}
\, .
\label{I1ak}
\ee

Combining  the bounds (\ref{IRj},\ref{I1ak}) on $G[x]$ with the estimate
\rf{I1ad} for $I_1(x)$ we finally arrive at the desired result \rf{I1b}
with
\be
C_1  \ =\ \left(\frac1{\pi(d-2)}\right)^{d-2}   \, ,
\quad \quad \quad \quad
C_2  \ =\ \sqrt{2} \, d^{5/4} \left(\frac12\right)^{d-2}   \, .
\label{Cs}
\ee

For $d=2$ it is not difficult to improve the lower bound
to $C_1 = \sqrt{2}$ while for
$d=3$ stricter estimates lead to the slightly stronger bounds
$C_1 = \sqrt{2}/\pi$, $C_2 = 3^{5/4} \sqrt{2}/\pi$.

\end{document}